\documentclass[12pt,usenames,dvipsnames,a4paper]{article}

\usepackage[utf8]{inputenc}
\usepackage[T1]{fontenc}

\usepackage{changepage}
\usepackage{amsmath,amsfonts,amssymb}
\usepackage{epsf,amsmath,bbold,amsfonts,stmaryrd}
\usepackage{mathrsfs}
\usepackage{appendix}
\usepackage{caption}
\usepackage{color}
\usepackage{datetime}
\usepackage{float}
\usepackage{graphicx}
\usepackage[colorlinks,hyperindex,breaklinks]{hyperref}
\hypersetup{pageanchor=false,citecolor=red,urlcolor=red}
\usepackage{indentfirst}
\usepackage[numbers,square,comma,sort&compress,merge]{natbib} 
\usepackage{subfig}
\usepackage{mathtools}
\usepackage{ytableau}
\usepackage{tabu}
\usepackage{esvect}

\usepackage{calc}
\usepackage{bm}

\allowdisplaybreaks

\def\c{\chi}

\def\d{\delta}

\def\eps{\varepsilon}
\def\f{\frac}
\def\g{\gamma}

\def\l{\left}

\def\mc{\mathcal}

\def\m{\mu}
\def\n{\nu}

\def\p{\partial}

\def\r{\right}
\def\s{\sigma}
\def\t{\tau}

\def\x{\xi}

\def\z{\zeta}

\def\be{\begin{equation}}
\def\ee{\end{equation}}

\def\bea{\begin{eqnarray}}
\def\eea{\end{eqnarray}}

\def\ba{\begin{array}}
\def\ea{\end{array}}

\def\bc{\begin{center}}
\def\ec{\end{center}}

\def\bl{\begin{flushleft}}
\def\el{\end{flushleft}}

\def\br{\begin{flushright}}
\def\er{\end{flushright}}

\def\bi{\begin{itemize}}
\def\ei{\end{itemize}}

\def\bt{\begin{tabular}}
\def\et{\end{tabular}}

\makeatletter

\newsavebox\myboxA
\newsavebox\myboxB
\newlength\mylenA

\newcommand*\xoverline[2][0.75]{%
    \sbox{\myboxA}{$\m@th#2$}%
    \setbox\myboxB\null
    \ht\myboxB=\ht\myboxA%
    \dp\myboxB=\dp\myboxA%
    \wd\myboxB=#1\wd\myboxA
    \sbox\myboxB{$\m@th\overline{\copy\myboxB}$}
    \setlength\mylenA{\the\wd\myboxA}
    \addtolength\mylenA{-\the\wd\myboxB}%
    \ifdim\wd\myboxB<\wd\myboxA%
       \rlap{\hskip 0.5\mylenA\usebox\myboxB}{\usebox\myboxA}%
    \else
        \hskip -0.5\mylenA\rlap{\usebox\myboxA}
         {\hskip 0.5\mylenA\usebox\myboxB}%
    \fi}
\makeatother

\def\be{\begin{equation}}
\def\ee{\end{equation}}

\def\bea{\begin{eqnarray}}
\def\eea{\end{eqnarray}}

\def\f{\frac}

\def\p{\partial}

\usepackage{xspace}

\usepackage{xifthen}
\newcommand*\diff{\mathrm{d}} 
\newcommand*\ldiff[2][]{ \ifthenelse{\isempty{#1}}{ \diff
#2}{\diff^#1#2} \,} 
\let\limitint\int 
\renewcommand{\int}{\limitint \!} 

\begin{document}

\begin{titlepage}

\vspace*{-1cm}

\begin{center}
\Large\textbf{Higgs inflation in Weyl-invariant Einstein-Cartan gravity} 
\end{center}

\begin{center}
\textsc{Georgios K. Karananas$^\star$ and Mikhail Shaposhnikov$^\dagger$}
\end{center}

\begin{center}
\it {$^\star$Arnold Sommerfeld Center\\
Ludwig-Maximilians-Universit\"at M\"unchen\\
Theresienstra{\ss}e 37, 80333 M\"unchen, Germany\\
\vspace{.4cm}
$^\dagger$Institute of Physics \\
\'Ecole Polytechnique F\'ed\'erale de Lausanne (EPFL) \\ 
CH-1015 Lausanne, Switzerland\\
\vspace{.4cm}
}
\end{center}

\begin{center}
\small
\texttt{\small georgios.karananas@physik.uni-muenchen.de}  \\
\texttt{\small mikhail.shaposhnikov@epfl.ch} 
\end{center}

\vspace*{2cm}

\begin{abstract} 

In this short note we analyze the inflationary dynamics in Weyl-invariant
Einstein-Cartan gravity coupled to the Standard Model of particle physics. We
take the axion-like particle of gravitational origin to be approximately
massless in the early Universe and show how inflation with the Higgs field
materializes.

\end{abstract}

\end{titlepage}

The inflationary dynamics of the Weyl-invariant Einstein-Cartan
gravity~\cite{Karananas:2024xja} with a heavy, decoupled, axion-like particle
(ALP) was discussed in length in~\cite{Gialamas:2024iyu,Karananas:2025fas}.

Here, we consider the full-blown theory: the ALP is taken to be a practically
massless spectator and thus present during inflation.\footnote{Note that
inflation with the ALP is also possible, as shown
in~\cite{Karananas:2025xcv}.} Provided that a sufficiently strong interaction
with QCD is generated, the field can eventually play the QCD axion
role~\cite{Karananas:2025ews}.

The action describing the Weyl-invariant Einstein-Cartan gravity has been
constructed in~\cite{Karananas:2024xja}, and reads
\begin{align}
\label{eq:infl_action_full}
S &= \int \diff^4 x \sqrt{g} 
\Bigg[
\f{1}{f^2}R^2 
+\f{1}{\tilde f^2}\tilde R^2
+ \f{\x_h h^2}{2} R 
+ \f{\z_h h^2}{2} \tilde R \nonumber \\
&~~\qquad\qquad+ \f{c_{aa}h^2}{2} a_\m a^\m 
+\f{c_{\t\t}h^2}{2} \t_{\m\n\rho}\t^{\m\n\rho}
+\f{\tilde c_{\t\t}h^2}{2}E^{\m\n\rho\s}\t_{\lambda\m\n}\t^\lambda_{\ \rho\s} 
-\f 1 2\l(D_\m^W h\r)^2 -\f{\lambda h^4}{4} 
\Bigg] \ ,
\end{align} 
where $g=-{\rm det}(g_{\m\n})$, $f,\tilde f$ are the gauge couplings of the
Lorentz group, $R$ and $\tilde R$ the scalar and pseudoscalar curvatures,
respectively, $\x_h,\z_h,c_{aa},c_{\t\t},\tilde c_{\t\t}$ are nonminimal
couplings of the Higgs $h$ (in unitary gauge) to the various gravitational
invariants comprising the curvatures as well as axial torsion $a_\m$ and the
reduced torsion tensor $\t_{\m\n\rho}$; we also introduced the densitized
totally antisymmetric symbol $E^{\m\n\rho\s} = \eps^{\m\n\rho\s}/\sqrt{g}$;
finally, $\lambda$ is the Higgs self-coupling, whereas the Weyl-covariant
derivative is defined as
\be
D_\m ^ W h = \p_\m h +\f{v_\m}{3} h \ ,
\ee
with $v_\m$ the torsion vector. 

The equivalent metric theory of~(\ref{eq:infl_action_full})---and how to
obtain it---was presented in~\cite{Karananas:2024xja,Karananas:2025fas}.
Briefly, one first expresses the action in its ``first-order'' form by
introducing two auxiliary fields, the spurion/dilaton $\c$ and the ALP
$\phi$, such that
\begin{align}
\label{eq:infl_action_full_first_order}
S &= \int \diff^4 x \sqrt{g} 
\Bigg[ 
\l(\c^2 + \f{\x_h h^2}{2}\r) R 
+\l(M_P^2\phi + \f{\z_h h^2}{2}\r)\tilde R 
-\f{f^2\c^2}{4} 
-\f{\tilde f^2 M_P^4\phi^2}{4}\nonumber \\
&~~\qquad\qquad+ \f{c_{aa}h^2}{2} a_\m a^\m 
+\f{c_{\t\t}h^2}{2} \t_{\m\n\rho}\t^{\m\n\rho} 
+\f{\tilde c_{\t\t}h^2}{2}E^{\m\n\rho\s}\t_{\lambda\m\n}\t^\lambda_{\ \rho\s} 
-\f 1 2\l(D_\m^W h\r)^2 
-\f{\lambda h^4}{4} 
\Bigg] \ .
\end{align} 
Then, one fixes the gauge by choosing $\c=M_P/\sqrt 2$, with $M_P$ the Planck
mass, resolves the curvature invariants into the Riemannian and torsional
pieces, and finally eliminates the Higgs-gravity coupling by redefining
appropriately the metric (see the aforementioned articles for details). At
the end of the day, the theory~(\ref{eq:infl_action_full}) is expressed as
\be
\label{eq:Einstein_frame_noncanonical}
S =\f 1 2 \int \diff^4x \sqrt{g} 
\Bigg[
M_P^2 \mathring R 
-\sum_{a,b}\widetilde\gamma_{ab} \p_\m \varphi^a \p^\m \varphi^b 
-\f{\lambda h^4}{2\Omega^4}
-\f{\tilde f^2M_P^2\phi^2}{2\Omega^4} 
-\f{f^2 M_P^4}{8\Omega^4} 
\Bigg] \ ,
\ee
where $\varphi^a=(h,\phi)$, $\widetilde \g_{ab}$ the metric of the
two-dimensional kinetic manifold, whose explicit expression is irrelevant for
the following and can be found in~\cite{Karananas:2024xja,Karananas:2025fas},
while $\Omega =\sqrt{1+\x_h h^2/M_P^2}$.

The kinetic sector can be diagonalized by introducing
\be
\label{eq:field_redef}
\Phi  = \f{M_P}{1+6\x_h}{\rm log}
\left(
\f{6+(1+6\x_h)\f{h^2}{M_P^2}}{\left| 3\z_h -(1+6\x_h)\phi \right|}
\right) \ ,
~~~H = \sqrt{6}M_P\, {\rm arctanh}
\left( 
\f{\f{h}{M_P}}{\sqrt{6+(1+6\x_h)\f{h^2}{M_P^2}}} 
\right) \ .
\ee
Notice that the original fields take values $h,\phi\in(-\infty,+\infty)$.
While this is still the case for $\Phi$, the nontrivial mapping between $H$
and $h$ translates into the former being defined in the compact domain $H\in
[-H_0,H_0]$, with $H_0 = \sqrt 6 M_P\, {\rm arccoth}\sqrt{1+6\x_h}$. Notice
also that possible nonanalyticities in~(\ref{eq:field_redef}) are
``coordinate'' singularities, as the scalar curvature of the kinetic manifold
remains regular at all field values.

In terms of $H$ and $\Phi$, the action~(\ref{eq:Einstein_frame_noncanonical})
becomes~\cite{Karananas:2024xja}
\be
\label{eq:canonical_Higgs_action}
S = \int\diff^4x\sqrt{g}
\Bigg[
\f{M_P^2}{2} \mathring R 
-\f 1 2(\p_\m H)^2
-\bar\g_{\Phi\Phi}(\p_\m\Phi)^2 
-V(H,\Phi) 
\Bigg] \ ,
\ee
with
\be
\bar\g_{\Phi\Phi} = \f 3 4 
\f{\cosh^4\l(\f{H}{\sqrt 6 M_P}\r)}{\cosh^2\l(\f{H}{\sqrt 6 M_P}\r)
\l(\f{1-\f{\z_h}{2}e^{(1+6\x_h)\f{\Phi}{M_P}}}{1+6\x_h}\r)^2
+\l(\f{e^{(1+6\x_h)\f{\Phi}{M_P}}}{24}\r)^2
\l(1+144c_{aa}\sinh^2\l(\f{H}{\sqrt 6 M_P}\r)\r)} \ ,
\ee
and
\begin{align}
\label{eq:potential_canonical_fields}
V(H,\Phi)& =  9\lambda M_P^4\sinh^4\l(\f{H}{\sqrt 6 M_P}\r)\nonumber \\
&\qquad
+\f{f^2M_P^4}{16}\l(1-6\x_h\sinh^2\l(\f{H}{\sqrt 6 M_P}\r)\r)^2\nonumber \\
&\qquad
+\f{9\tilde f^2 M_P^4 e^{-2(1+6\x_h)\f{\Phi}{M_P}}}{(1+6\x_h)^2}
\Bigg[
\cosh^2\l(\f{H}{\sqrt 6 M_P}\r)\nonumber \\
&\qquad\qquad\qquad
-\f{\z_h}{2} e^{(1+6\x_h)\f{\Phi}{M_P}}
\l(1-6\x_h\sinh^2\l(\f{H}{\sqrt 6 M_P}\r)\r)
\Bigg]^2\ .
\end{align}

There are some comments that we wish to make here. The first concerns the
Lorentz group gauge couplings $f$ and $\tilde f$ that in the equivalent
metric theory appear in the potential, see the second and third lines of the
above Eq.~(\ref{eq:potential_canonical_fields}). In our setup, the former
sets the classical value of the cosmological constant $\Lambda$, as well as
the gravitationally-induced (classical) value of the Higgs mass $m_{h,{\rm
grav}}$. The latter controls the gravitationally-induced ALP mass $m_
{\Phi,{\rm grav}}$. Indeed,
expanding~(\ref{eq:canonical_Higgs_action})-(\ref{eq:potential_canonical_fields})
around the minimum of the potential which is located at
\be
H=0,~~~\Phi=\f{M_P}{1+6\xi_h}\log\f{2}{|\z_h|} \ , 
\ee
we find (after canonically normalizing $\Phi$)
\be
\Lambda = \f{f^2M_P^2}{16} \ ,
~~~m_{h,{\rm grav}}^2 = - \f{f^2\x_h M_P^2}{4} \ ,
~~~m_{\Phi,{\rm grav}}^2 = \f{\tilde f^2(1+6\x_h)M_P^2}{48} \ .
\ee
For the correct hierarchy between $\Lambda$ and $M_P$ to be achieved we need
to require that $f\lll 1$, which also forces $m_{h,{\rm grav}}$ to be
practically zero. At the same time, for the strong-CP puzzle to be addressed
in this construction, $m_{\Phi,{\rm grav}}$ needs to be smaller than the
QCD-induced mass, which in turn means that $\tilde f \lll 1$. Remembering
that the origin of the couplings $f$ and $\tilde f$ is gravitational, we find
it logical that these be tiny. On the other hand, the selfconsistency of the
theory and more specifically the absence of accidental gauged
symmetries~\cite {Karananas:2024hoh,Karananas:2024qrz}\,\footnote{See
also~\cite{Alvarez-Gaume:2015rwa,Golovnev:2023zen,Hell:2023mph,Casado-Turrion:2024esi,Hell:2025wha,Barker:2025gon}~for
related considerations.}~forces both of them to be strictly nonvanishing,
irrespectively of their smallness.

Our second comment is that, as is, the
theory~(\ref{eq:canonical_Higgs_action}) cannot Higgs-inflate the Universe.
The discussion in the previous paragraph makes it clear that the most
interesting situation is when the $\lambda$-part of the potential---given in
the first line of~(\ref {eq:potential_canonical_fields})---dominates. Being a
hyperbolic sine, it can lead to inflation of quartic selfinteraction type,
whose predictions are ruled out by observational
data~\cite{Planck:2018jri,BICEP:2021xfz}.

Another comment concerns the complete nonappearance of the parameters $\x_h$,
$c_ {aa}$ and $\z_h$ in the Higgs sector of the tree-level action, in
contradistinction with the decoupled-ALP
case---see~\cite{Karananas:2025fas}---where the latter two shape the
inflationary behavior of the Higgs. We notice
from~(\ref{eq:canonical_Higgs_action})-(\ref{eq:potential_canonical_fields})
that these feed exclusively into the ALP-sector. Therefore, even if chosen in
one way or another, the conclusion about the non-existence of inflationary
dynamics, classically, remains true.

And as a final comment, notice that the Higgs dynamics is completely
unaffected by the ALP: the field has a canonical kinetic term and its
potential is dominated by the only term not involving $\Phi$. As we point out
now this has important consequences.

Let us assume that $H\ll \sqrt{6} M_P$, so that we are well within the
validity domain of the theory; then, to leading order in the Higgs field, the
parts of the action that comprise only $H$ read
\be
\label{eq:minimally_coupled_H}
S \approx \int\diff^4x \sqrt{g} 
\l[
\f{M_P^2}{2}\mathring R 
-\f 1 2 (\p_\m H)^2 
-\f{\lambda}{4}H^4  
\r] \ . 
\ee
Surprisingly, up to Planck-suppressed operators, it all boils down to the
textbook situation of a (minimally coupled to gravity) canonical scalar with
quartic potential.  Due to the nonrenormalizability of the theory, a plethora
of operators will be generated at the quantum level. Among those, and
provided that loop effects are accounted for in the metric-equivalent
form~(\ref{eq:minimally_coupled_H}) of the theory, there is a unique
dimension-four contribution: a nonminimal interaction between the canonical
Higgs field and scalar curvature, so as to tame ultraviolet divergences
$\propto H^2\mathring R$. Its presence is therefore an inevitable requirement
for quantizing the theory in arbitrary backgrounds.  Accounting for that, the
selfconsistency of the metric-equivalent theory dictates
that~(\ref{eq:minimally_coupled_H}) be extended as
\be
\label{eq:nonminimally_coupled_H}
S = \int \diff^4x \sqrt{g}
\l[
\f{M_P^2 +\widetilde{\x}_H H^2}{2}\mathring R 
-\f 1 2 (\p_\m H)^2 
-\f{\lambda}{4}H^4  
\r] \ ,
\ee
which is exactly the action of the original Higgs
inflation~\cite{Bezrukov:2009db}. Due to the presence of the nonminimal
coupling $\widetilde \x_H$, the state of affairs can  radically change. Its
value is a priori arbitrary - remember, it is an undetermined parameter of
the effective theory. If $\widetilde \x_H$ is taken to be small, it is well
known that there is no successful inflation. If on the other hand
$\widetilde{\x}_H\sim \mc O(10^{3-4})$, the potential flattens, slow-roll
occurs for $\f{M_P}{\sqrt{\widetilde{\x}_H}}\ll H\ll
\sqrt 6 M_P$, and the inflationary predictions are in excellent agreement
with the Planck/BICEP observations~\cite{Planck:2018jri,BICEP:2021xfz}.
Inevitably, a large nonminimal coupling lowers (significantly) the scale of
perturbative unitarity, and hence tightens the energy range over which the
theory is applicable; see
eg.~\cite{Burgess:2009ea,Barbon:2009ya,Burgess:2010zq,Bezrukov:2010jz,Karananas:2022byw}.

One may wonder why loop-induced contributions to the nonminimal coupling are
not explicitly presented for ``conventional'' Higgs inflation in the
so-called Einstein frame, both in the metric~\cite{Bezrukov:2007ep} and the
Palatini formalisms~\cite{Bauer:2008zj,Shaposhnikov:2020fdv}. Of course, such
corrections can and will be generated; however, they are exponentially
small~\cite{Bezrukov:2010jz}, due to an approximate shift symmetry for the
inflaton kicking-in at large field values.

Importantly, if  one instead works in the EC formulation, terms involving
torsional invariants can in principle appear. In the metric-equivalent theory
these manifest themselves as novel interactions between and among the Higgs,
as well as gravitational degrees of freedom, i.e. the curvature and
ALP.\footnote{A more detailed study of higher-dimensional operators as well
as their effects on inflation is left for future
work~\cite{ToAppear_higher_dimensions}.} For instance, the nonmiminal
coupling in the metrical theory~(\ref{eq:nonminimally_coupled_H}) corresponds
to modifying~(\ref{eq:infl_action_full_first_order}) as
\be
S \mapsto S + \d S \ ,
~~~\d S =  \int\diff^4 x \sqrt g 
\l[
c_1 h^2\f{\l(D_\m^W \c\r)^2}{\c^2} + c_2 D^{W\m} h^2\f{D_\m^W\c}{\c}
\r] \ ,
\ee
with 
\be
c_1 =- \f 2 9(1+6\widetilde \x_H)\ ,
~~~c_2 =\f 1 6 (1+6\widetilde \x_H)\ ,
~~~D_\m ^ W \chi = \p_\m \chi +\f{v_\m}{3} \chi \ .
\ee
Notice that both terms in the above are manifestly Weyl-invariant.

Before closing, let us stress that a crucial open question concerns the
light-ALP dynamics during and after inflation. Answering it requires a
separate investigation and is left for the forthcoming
article~\cite{ToAppear}.

\begin{center}
\textbf{Acknowledgements} 
\end{center} 

We thank Sebastian Zell for many fruitful discussions and comments on the
manuscript.

\begin{center}
\textbf{References} 
\end{center}

{

\small

\setlength\bibsep{0pt}
    \bibliographystyle{utphys}
 \bibliography{Refs}
}

\end{document}